\documentclass[conference]{IEEEtran}
\IEEEoverridecommandlockouts
\usepackage{cite}
\usepackage{amsmath,amssymb,amsfonts}
\usepackage{braket}
\usepackage{algorithmic}
\usepackage{graphicx}
\usepackage{textcomp}
\usepackage{hyperref}
\usepackage{xcolor}
\def\BibTeX{{\rm B\kern-.05em{\sc i\kern-.025em b}\kern-.08em
    T\kern-.1667em\lower.7ex\hbox{E}\kern-.125emX}}

\usepackage[T1]{fontenc}

\usepackage{listingsutf8}
\definecolor{codegreen}{rgb}{0,0.6,0}
\definecolor{codegray}{rgb}{0.5,0.5,0.5}
\definecolor{codepurple}{rgb}{0.58,0,0.82}
\definecolor{backcolour}{rgb}{0.95, 0.95, 0.95}

\usepackage{orcidlink}

\lstdefinestyle{codeblock}{
  backgroundcolor=\color{backcolour},
  commentstyle=\color{codegreen},
  keywordstyle=\color{blue},
  numberstyle=\tiny\color{codegray},
  stringstyle=\color{codepurple},
  basicstyle=\footnotesize,
  escapechar=\¢, 
  otherkeywords={with},
  breakatwhitespace=false,
  breaklines=true,
  captionpos=b,
  keepspaces=true,
  language=Python,
  numbers=left,
  numbersep=5pt,
  showspaces=false,
  showstringspaces=false,
  showtabs=false,
  tabsize=2,
  basicstyle=\ttfamily\footnotesize,
  inputencoding=utf8,
  upquote=true,
}
\makeatletter
\newcommand{\linebreakand}{%
  \end{@IEEEauthorhalign}
  \hfill\mbox{}\par
  \mbox{}\hfill\begin{@IEEEauthorhalign}
}
\makeatother

\begin{document}

\title{Constant-time hybrid compilation of Shor's algorithm with quantum
  just-in-time compilation}

\author{\IEEEauthorblockN{David Ittah\IEEEauthorrefmark{1} \orcidlink{0000-0003-0975-6448}, Jackson Fraser\IEEEauthorrefmark{2}, Josh Izaac\IEEEauthorrefmark{1} \orcidlink{0000-0003-2640-0734}, and Olivia Di Matteo\IEEEauthorrefmark{3} \orcidlink{0000-0002-1372-7706}}
\IEEEauthorblockA{\IEEEauthorrefmark{1}\textit{Xanadu}, Toronto, ON, M5G 2C8, Canada}
\IEEEauthorblockA{\IEEEauthorrefmark{2}\textit{Department of Physics and Astronomy, The University of British Columbia}, Vancouver, BC, Canada \\
Email: jacksonfraser55@gmail.com}
\IEEEauthorblockA{\IEEEauthorrefmark{3}\textit{Department of Electrical and Computer Engineering and Stewart Blusson Quantum Matter Institute}, \\\textit{The University of British Columbia}, Vancouver, BC, Canada \\
Email: olivia.dimatteo@ubc.ca}}

\maketitle

\begin{abstract}
Continuous improvements in quantum computing hardware are exposing the need for simultaneous advances in software. Large-scale implementation of quantum algorithms requires rapid and automated compilation routines such as circuit synthesis and optimization. 
As systems move towards fault-tolerance, programming frameworks and compilers must also be capable of compiling and optimizing programs comprising both classical and quantum code. 
This work takes a step in that direction by providing an implementation of Shor's factoring algorithm, compiled to elementary quantum gates using PennyLane and Catalyst, a library for quantum just-in-time (QJIT) compilation of hybrid workflows. 
We demonstrate that with QJIT compilation, the algorithm is compiled once per \emph{bit width} of $N$, the integer being factored, even when $N$-specific optimizations are applied to circuit generation based on values determined at runtime. 
The implementation is benchmarked up to 32-bit $N$, and both the
size of the compiled program and the pure compilation time are found to be constant (under 3 seconds on a laptop computer), meaning code generation becomes tractable even for realistic problem sizes.
\end{abstract}


\section{Introduction} 

Quantum computing, if realized at scale, promises exponential scaling advantages over the best-known classical algorithms for some important real-world problems.
However, actually implementing quantum algorithms for problem sizes of interest is an outstanding challenge.
Most quantum programming is done (and taught) in the circuit model, a low-level abstraction where a programmer specifies how individual quantum gates are applied to qubits.
Improvements in hardware, in particular the growing software needs of error correction, necessitate novel abstractions that will narrow the gap and make productive quantum programming as accessible as classical programming
\cite{di2024abstraction}.

One key approach to narrowing this gap is improvements in quantum compilation. A programmer should be able to express algorithms using a combination of classical and quantum functions, and rely on a compiler to automatically and effectively decompose and optimize that program into low-level instructions. 
While most programming frameworks are equipped with quantum compilation tools to address optimization of circuits, they neglect the surrounding classical code and its interaction with the quantum portion.
Addressing such \emph{hybrid} compilation is essential, as all quantum algorithms, in the context of solving a real-world problem, necessitate exchange between a classical and quantum processor. 

A canonical example is Shor's algorithm.
A full implementation involves the interaction of many moving parts, both classical and quantum.
Quantum circuits for modular exponentiation, the core of the quantum subroutine, must be developed, decomposed, and optimized.
Though quantum programming frameworks generally contain the constituent subroutines (e.g., the quantum Fourier transform (QFT), adder, and multiplier circuits), their specific implementations may not be conducive for Shor's algorithm (for instance, a built-in QFT may swap the order of qubits at the end, while other subroutines assume the opposite qubit ordering).
Classical utility functions are also required to process measurement outcomes, such as estimating the numerator and denominator of a floating point number.

Nevertheless, there are some notable recent implementations. 
Ref.~\cite{peng2022formally} develops a formally-verified implementation of both the classical and quantum subroutines using the Coq proof assistant. Circuits for problem instances up to 10-bit integers are expressed in the low-level SQIR language, from which OpenQASM code is extracted. To that end, the classical and quantum portions are extracted separately. 

Working with the higher-level language Qrisp, Ref.~\cite{polimeni25_end_to_end_compil_implem} recently demonstrated end-to-end compilation of Shor's algorithm for the discrete logarithm problem, with application to elliptic curve cryptography. With Qrisp, subroutines applied to registers of qubits are expressed naturally using functions and arithmetic applied to quantum data types. Ref.~\cite{polimeni25_end_to_end_compil_implem} focuses on the traditional quantum compilation aspects (circuit generation and optimization). It does not consider the surrounding classical control flow, but acknowledges this as a next step in the concluding remarks. 
In a similar vein, the Classiq library \cite{classiq-library2024} includes an implementation in the high-level language Qmod \cite{qmod}, using the same circuits as the present work; however the quantum component is generated  and optimized separately from the surrounding classical control flow.

The key contribution of this work is to demonstrate how quantum just-in-time (QJIT) compilation can be used to compile the \emph{entire} workflow of Shor's algorithm, both classical and quantum portions, together. Catalyst \cite{Ittah2024} is used to compile code written using Python and PennyLane \cite{bergholm2022pennylaneautomaticdifferentiationhybrid} to an intermediate representation comprising both classical instructions and elementary quantum gates. Remarkably, with the right choice of data types and control flow, QJIT compilation takes \emph{constant} time, independent of the number being factored, even when optimizations are applied based on random parameters chosen at runtime. On a modest laptop, this takes under 3 seconds, even for the largest problem size tested (a 32-bit number, which would decompose into roughly 40 million 1- and 2-qubit gates at runtime). Moreover, recompilation is avoided when factoring different integers with the same bit width, highlighting the utility of QJIT compilation for algorithms at scale. A full implementation, including benchmarking scripts, is available at \url{https://github.com/QSAR-UBC/shortalyst}.

Below, Sec.~\ref{sec:background} provides background information about QJIT compilation and Shor's algorithm. Sec.~\ref{sec:methods} outlines the implementation, including choice of quantum circuits, optimizations, and special considerations to enable QJIT compilation. Benchmarking results are presented in Sec.~\ref{sec:results}, followed by concluding remarks in Sec.~\ref{sec:conclusions}.

\section{Background}
\label{sec:background}

\subsection{Quantum just-in-time compilation}

Quantum software frameworks, such as Qiskit \cite{qiskit}, PennyLane, and Cirq \cite{cirq} (to name a few), have been invaluable tools for quantum algorithm exploration and development over the past 5 years. Predominantly Python-based, these tools allow for a mixing of classical processing (in Python) alongside quantum instruction sets, but with certain restrictions:

\begin{itemize}
    \item Classical processing and control flow may be used to generate quantum instructions --- as long as the information needed is statically available during Python runtime (that is, not outputs of quantum processing).
    \item Once classical processing is complete, a flat, unrolled list of quantum instructions is sent for processing via a quantum simulator or hardware device, and the result returned to Python.
\end{itemize}

However, it is becoming clear that this separation of classical and quantum processing is not ideal, for several reasons. First, it does not allow for easy incorporation of dynamic information, e.g., mid-circuit measurement outcomes, in the definition of algorithms (a requirement in workflows like iterative quantum phase estimation or quantum error correction). More importantly, however, is that quantum compilation must be repeated every time quantum instructions are generated, all while the list of instructions will grow \textit{increasingly large} as quantum algorithms scale. Instead, what is needed is the ability to compile hybrid workflows \textit{directly}. This would preserve the structure of the original algorithm outside of Python, reducing the need for repeated quantum compilation.

Just-in-time (JIT) compilation is a solution to this problem. JIT compilation first appeared in the Python machine-learning ecosystem in tools such as Numba \cite{numba}, PyTorch \cite{pytorch}, and JAX \cite{jax2018github}. In this approach, algorithms and workflows continue to be written in Python---in some cases utilizing native Python control flow---however, they are not executed by the Python interpreter. Instead, when Python functions are called by the user, the structure and contents of the function are captured and compiled to optimized machine code, which is then executed instead. On subsequent calls from Python, the pre-compiled binary can directly be executed, leading to significant performance improvements (both from having staged computation out of Python, and only paying the price of compilation once).

This work utilizes Catalyst \cite{Ittah2024}, a quantum just-in-time (QJIT) compiler for PennyLane Python programs. Inspired by the solutions above, Catalyst captures the entire hybrid workflow for efficient compilation and staged-out execution.

\begin{enumerate}
    \item First, the workflow must be coded using PennyLane (for quantum instructions) and JAX (for classical instructions and linear algebra). The entry-point to this workflow must then be marked with the \texttt{@qjit} decorator.
    \item When the workflow is first invoked by the user, the \texttt{@qjit}-decorated code is captured by Catalyst, including classical processing, native Python control flow (using the AutoGraph feature), and quantum instructions. The program is then analyzed and transformed on a rich hybrid intermediate representation (IR) in the MLIR framework \cite{mlir}, and lowered to low-level instructions with the LLVM compiler toolchain~\cite{llvm} to produce an optimized, parametrized binary. During this process, both classical and quantum optimizations are applied, while preserving the algorithmic structure (in particular loops, subroutines, and dynamic decision structures).
    \item Once compilation is complete, the parametrized binary will be executed on the specified quantum simulator or hardware backend with the provided runtime parameters.
\end{enumerate}

Importantly, the next time the workflow is executed, compilation (step 2) is not repeated, even if runtime parameters are modified --- provided the input type signature stays the same.

\subsection{Shor's algorithm}
\label{subsec:shor_overview}

Shor's algorithm \cite{shor1997} is a prime example of an algorithm that stands to benefit from QJIT compilation. The input to the algorithm is an integer, $N$, promised to be a product of two prime integers $p$ and $q$, with bit width $n = \lfloor \log_2 N \rfloor + 1$. The goal is to recover $p$ and $q$ such that $N = pq$. 

Factoring with Shor's algorithm is accomplished by a reduction to \emph{order-finding} \cite{miller1976riemann, shor1997}. 
The order of an integer $a$ modulo $N$ is the smallest non-zero integer, $r$, for which $a^r \equiv 1 \pmod N$. 
Shor's algorithm begins by randomly selecting $a$ from the range $2 \leq a < N - 1$. 
If $N$ and $a$ have no common factor, a quantum computer is used to determine a candidate for the order of $a$ modulo $N$. 
If the candidate order, $r$, is even, then $r = 2r^\prime$ and the definition of the order reduces to 
\begin{equation}
a^{2r^\prime} - 1 = (a^{r^\prime} - 1)(a^{r^\prime} + 1)  \equiv 0 \pmod N.
\label{eq:non-trivial-sq-root}
\end{equation}

\noindent If $a^\prime = a^{r^\prime}$ is a non-trivial square root, i.e., $a^\prime \not\equiv 1 \pmod N$ and $a^\prime \not\equiv N - 1 \pmod N$, then Eq.~\ref{eq:non-trivial-sq-root} indicates that $a^{r^\prime} \pm 1$ are each a multiple of either $p$ or $q$. 
Thus, $p$ and $q$ can be recovered by computing the greatest common divisors of $a^{r^\prime} \pm 1$ with $N$. 
Shor's algorithm, as described, succeeds with probability 1/2 \cite{miller1976riemann, shor1997}.
If the square root is trivial, a new $a$ can be chosen and the procedure repeated. 

Pseudocode for Shor's algorithm is presented below.

\begin{lstlisting}
def shors_algorithm(N):
    p, q = 0, 0

    while p * q != N:
        a = random.choice(range(2, N - 1))

        if gcd(N, a) != 1:
            p = gcd(N, a)
            return p, N // p

        cand_r = find_candidate_order(N, a)

        if cand_r % 2 == 0:
            root = (a ** (cand_r // 2)) % N

            if root not in [1, N - 1]:
                p = gcd(N, root - 1)
                q = gcd(N, root + 1)

    return p, q  
\end{lstlisting}

Expressed in this way, the quantum computing aspects are entirely abstracted away in the \texttt{find\_candidate\_order} subroutine.
Shor's algorithm can be disguised, in a sense, as a classical algorithm.
Of course, any practical implementation must be compiled to executable code, and this is where complications arise.

The \texttt{find\_candidate\_order} subroutine involves construction of quantum circuits based on the values of $N$ and $a$, optimization of these circuits, execution on the device (hardware or simulator), followed by post-processing of results.
The key complication is that the value of $a$ is selected randomly \emph{at runtime}. 
With an interpreted language, and without prior knowledge of $a$, circuits must either (a) also be generated and optimized at runtime, or (b) generated and optimized in advance, stored, and read from memory at runtime.
Shor's algorithm has a success probability of 1/2, so if more than one $a$ must be tried, the first option will lead to additional compilation time. 
However, even if it succeeds on the first attempt, that compilation overhead will still be incurred for new values of $N$. In that case, similar issues arise, as previously-stored circuits cannot be reused for multiple $N$.

\begin{figure*}[ht]
\centerline{\includegraphics[width=0.95\textwidth]{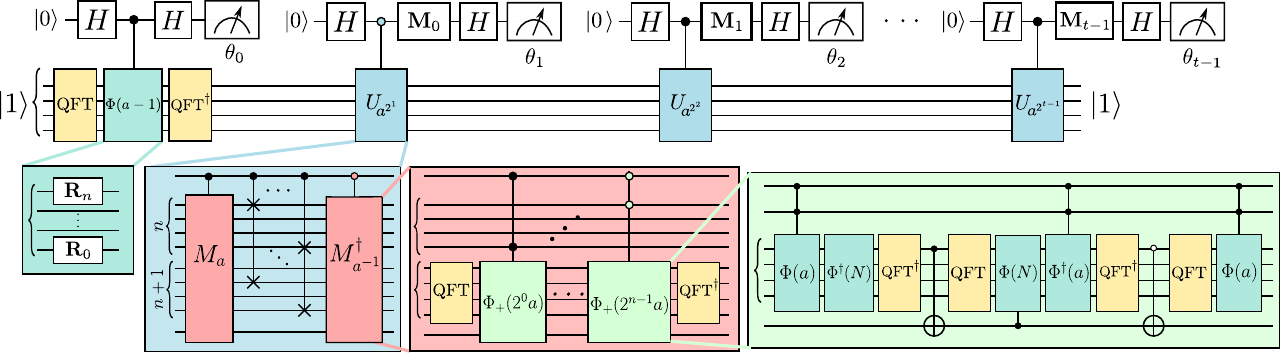}}
\caption{High-level structure of the QPE circuit in Shor's algorithm. This decomposition is based on the work of Beauregard \cite{beauregard2003circuits}.
  Insets show decompositions of subroutines. The operation $M_a$ performs $M_a (\ket{x}\ket{b}\ket{0}) = \ket{x}\ket{(b+ax)\pmod N} \ket{0}$. $\Phi$ and $\Phi_+$ perform addition in the Fourier basis \cite{draper2000additionquantumcomputer} (regular, and modulo $N$, respectively). $\mathbf{R}_k$ adds phase based on the bits of $a$, $\mathbf{R}_k = \hbox{diag}\left[(1, \exp\left(2\pi i \sum_{\ell =0}^k a_\ell / 2^{\ell +1}\right) \right]$, where $a = \sum_{k=0}^{n-1} 2^k a_k$. $\mathbf{M}_k$ adds phase based on measurement outcomes, $\mathbf{M}_k = \hbox{diag}\left[1, \exp\left(-2\pi i\sum_{\ell=0}^{k} \theta_{\ell} / 2^{k + 2 - \ell}\right)\right]$. Image adapted from Ref.~\cite{OliviaDiMatteo2025}. }
\label{fig:qpe_full_combined}
\end{figure*}

\section{QJIT-compiled implementation of Shor's algorithm}
\label{sec:methods}

The issues in Sec.~\ref{subsec:shor_overview} can both be resolved with QJIT compilation.
This section presents the quantum circuit construction, opportunities for optimization at runtime, and also outlines several caveats. 
In particular, the presented implementation requires the bit width of $N$ is passed as a static argument to compilation\footnote{We expect this limitation will be resolved in a future version of Catalyst.} 
The full implementation is available open-source on GitHub\footnote{\url{https://github.com/QSAR-UBC/shortalyst}}.

\subsection{Quantum circuit construction}

The \texttt{find\_candidate\_order} subroutine performs quantum phase estimation (QPE) for the unitary $U_a$ \cite{shor1997, nielsenandchuang},
\begin{equation}
    U_a \ket{x} = \ket{ax \pmod N}.
\end{equation}
\noindent Applying $U_a$ to a register in $\ket{1}$ a total of $j$ times yields $(U_a)^j \ket{1} = \ket{a^j \pmod N}$. Thus, after successive application of powers of $(U_a)^{2^k}$ in QPE, controlled on a $t$-qubit estimation register in uniform superposition, the combined state is \cite{nielsenandchuang}
\begin{equation}
    \left( \sum_{j = 0}^{2^t - 1} \ket{j} \right) \ket{1} \rightarrow \sum_{j = 0}^{2^t - 1} \ket{j} \ket{a^j \pmod N}.
\end{equation}
A key feature of this state is that the target register contains a superposition of possible powers of $a$. In Sec.~\ref{subsec:a-specific-opts}, knowledge of these powers is leveraged to perform $a$-specific circuit optimization at runtime. 

The chosen implementation for $U_a$ is that of Beauregard
\cite{beauregard2003circuits}. For an $n$-bit integer $N$, QPE uses $2n + 3$ qubits, where
\begin{itemize}
\item $n$ are in the target register,
\item $1$ is an estimation qubit, and
\item $n + 2$ are auxiliary qubits for modular exponentiation.
\end{itemize}

The implementation of the individual $U_a$ comprises a series of controlled multiplications modulo $N$, each of which decompose into a series of doubly-controlled additions in the Fourier basis. The high-level structure of the circuit is presented in Fig.~\ref{fig:qpe_full_combined}; we refer the reader to Ref.~\cite{beauregard2003circuits} and to the software implementation for further details.

The single estimation qubit is achieved using the well-known semiclassical version of the quantum Fourier transform \cite{griffiths1996semiclassical, parker2000efficient, beauregard2003circuits}. After each controlled operation, the estimation qubit is adjusted based on previous measurement outcomes, then measured. Note that this assumes the ability to feed measurement outcomes forward during execution, and dynamically adjust the circuit.
Such adjustment on the fly adds complexity in the representation of quantum programs (IRs), and especially compilation routines, as analyses and optimizations may need to be extended to account for program structure \cite{ittah2022, amy2025}. Decision making within the compiler can also be made more difficult, for example when setting optimization schedules (when should a subroutine be inlined, or abstractions lowered, with respect to optimization passes).

\subsection{Instance-specific optimizations}
\label{subsec:a-specific-opts}

We note that the circuits in Ref.~\cite{beauregard2003circuits} are not the most
optimal circuits, and that many implementations with better quantum resource counts exist (see Ref.~\cite{wang2024comprehensivestudyquantumarithmetic} for a recent and comprehensive review of arithmetic circuits). 
The goal of this work is not to provide the most optimal quantum circuit, but rather, to demonstrate that by using QJIT compilation, optimizations based on runtime data (i.e., differences in circuit construction that depend on $N$ and $a$) can be applied \emph{without triggering recompilation}. 
Another notable feature of the circuits in Ref.~\cite{beauregard2003circuits} is that their structure exposes opportunities for circuit optimization at higher abstraction levels than gates, i.e., at the level of entire subroutines.

Three main $a$-specific optimizations are implemented. The first is well-known: the powers $a^{2^k}$ can be precomputed so $U_{a^{2^k}}$ is applied in lieu of $(U_{a})^{2^k}$. This applies also when $a$ is chosen at runtime, reducing the number of controlled operations from $2^t - 1$ to $t$, where $t$ was the original number of estimation qubits. Furthermore, since the first $U_a$ is applied to a target register in $\ket{1}$, controlled addition of $a - 1$ in the Fourier basis can be used instead (in fact, this was leveraged to save resources in the hand-optimized circuits of the first experimental demonstration of factoring $N=15$ \cite{vandersypen2001experimental}).

The other two optimizations pertain to conditional operations. Similar routines were described in terms of bit operations in Ref.~\cite{beckman1996efficient}, but the analogous quantum implementation was expressed with controlled operations that always execute. We emphasize that here, QJIT compilation allows  \emph{removal} of entire portions of the circuit dynamically, at runtime. 

\begin{figure}[htbp]
\centerline{\includegraphics[width=0.93\columnwidth]{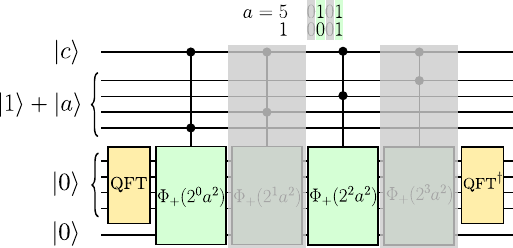}}
\caption[]{Instance-specific optimization enabling removal of doubly-controlled adders in the $M_a$ subroutine. Using the knowledge that the only basis states input to the control register are $\ket{1}$ and $\ket{a}$, we observe that the only doubly-controlled adders which get triggered are the first one, and those controlled on qubits that are $1$ in the binary representation of $a$ (more generally, the bitwise logical OR of possible powers of $a$). Image adapted from Ref.~\cite{OliviaDiMatteo2025}.}
\label{fig:bitwise-optim}
\end{figure}

The second optimization leverages knowledge of $a$ to eliminate redundant doubly-controlled adders in the controlled multiplication ($M_a$) subroutine. As illustrated in Fig.~\ref{fig:bitwise-optim}, by examining the superposition in the control register, certain doubly-controlled adders can be omitted entirely when the logical OR of the bits in the control register evaluates to zero.

\begin{figure}[ht]
\centerline{\includegraphics[width=0.93\columnwidth]{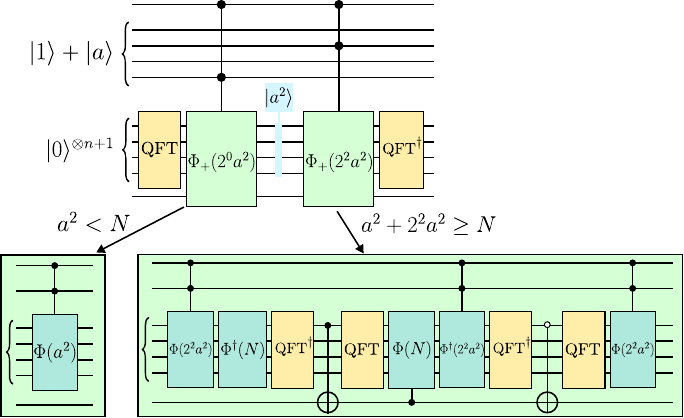}}
\caption[]{Instance-specific optimization relating to the overflow check and correction circuit in the Fourier adder modulo $N$. Since the register being added to starts in $\ket{0}$, the overflow check and correction circuit on the bottom right only becomes necessary when the terms in the superposition could lead to overflow (i.e., when their sum exceeds $N$). Image adapted from Ref.~\cite{OliviaDiMatteo2025}.}
\label{fig:skip-oflow-optim}
\end{figure}

The third $a$-specific optimization  removes the sometimes unnecessary overflow detection and correction circuit in Fourier addition modulo $N$. Fig.~\ref{fig:skip-oflow-optim} provides a graphical example. In the standard implementation, several gates are required to check and correct overflow conditions to ensure the addition is modulo $N$. However, when the value of $a$ is known, the overflow check and correction circuit can be removed until there are terms in the superposition where addition (or subtraction) leads to values above (or below) $N$. For the Fourier adders in $M_a$ this is straightforward, as the register being added to always starts in $\ket{0}$. The analogous process for $M^{\dagger}_{a^{-1}}$ requires additional bookkeeping to keep track of the terms in superposition.
Eventually we expect diminishing returns, as each controlled multiplication adds more terms into the superposition. 
Sec.~\ref{sec:results} quantifies the compilation overhead and quantum gate savings introduced by these optimizations.

\subsection{Quantum just-in-time compiling Shor's algorithm}

Going from a traditional Python-based implementation of Shor's algorithm to a scalable QJIT-compiled version requires:
\begin{enumerate}
    \item staging classical computation out of Python,
    \item capturing program structure, in particular control flow,
    \item and ensuring IR size does not grow with problem size.
\end{enumerate}

The following sections describe each requirement and how it is achieved using Catalyst.

\subsubsection{Classical computation}

While readers should be familiar with the concept of staging out quantum computation from Python (a process performed by most quantum SDKs when generating a circuit), the same can also be done with classical computation. For this purpose, Catalyst uses JAX as a frontend to capture classical computation expressed through Python arithmetic and the widely-adopted NumPy API \cite{numpy}. To capture computation, JAX evaluates Python functions on symbolic arguments, or \emph{abstract values}, that carry type information and record the sequence of operations applied to them in a global context---a process known in JAX as \emph{tracing} \cite{tracing}. 

Staging out computation in this way can result in additional restrictions (see \cite{jax-sharp-bits, catalyst-sharp-bits}) that must be respected by the source program. In particular, the lack of concrete values for program variables requires a certain code style that favours mathematical expressions over otherwise Pythonic constructs, vectorized operations over loops and generators, and pre-allocated result buffers over accumulation in a list. 

Despite such restrictions, this technique enables Catalyst to embed a wide range of computation side-by-side with quantum instructions into the hybrid program, as classical pre-processing, post-processing, or even intermediate processing on quantum measurements. The availability of the NumPy API plays a key role in being able to express complex arithmetic performed by algorithms such as Shor's, and enables ($a$-specific) optimizations like those described in Sec.~\ref{subsec:a-specific-opts}. Classical computation is also key to the parametrizability of the compiled program; by capturing the mapping from the high-level algorithmic inputs to the internal program parameters, users can re-run their program with new inputs without necessarily recompiling it.

\subsubsection{Control flow} Even more so than classical computation, capturing control flow may have the largest impact on both the expressibility and scalability of quantum workflows. Expressibility is improved because runtime-based control flow enables entirely new paradigms like repeat-until-success loops and adaptive gate application. Better scalability is achieved because loops and subroutines in particular can compress the program representation by several orders of magnitude. Consequently, everything from IR generation, to optimizations, to code generation is accelerated significantly.

For this purpose, Catalyst provides a functional form of common control flow routines compatible with the tracing mechanism: \texttt{for} loops, \texttt{while} loops, and \texttt{if-elif-else} branches \cite{catalyst-control-flow}. These act as specially-controlled subroutines with their own tracing scope, and translate to structured control flow operations in Catalyst's IR. While effective, this form of control flow is not the most user-friendly, and adds additional verbosity. To solve this problem, a library for Python operator overloading is used, called \textit{AutoGraph} \cite{autograph, catalyst-autograph}. By directly operating on the Python AST of targeted functions, language keywords like \texttt{if} and \texttt{for} can be replaced with calls to custom implementations of these expressions, allowing Catalyst to automatically dispatch to their functional forms. In this way, Shor's algorithm can be implemented in a JIT-friendly manner while retaining a high degree of readability.

\subsubsection{Constant IR size} While capturing classical computation and program structure should achieve a constant IR size, there were some additional pitfalls. Library subroutines, such as from PennyLane, may not be tracing-friendly or affected by AutoGraph, and so had to be written from scratch (e.g., multipliers, adders, and the QFT). Additionally, while generator expressions and list comprehensions are quite natural in Python, since these are executed by the interpreter they result in a large number of additional instructions in the IR. A better approach is to generate the elements on the fly in existing loop structures, or make use of the \texttt{vmap} operator to efficiently repeat a computation over different inputs.


A remaining limitation in the QJIT compilation process is the use of dynamically-shaped arrays, i.e., arrays whose length (or shape) is not known until runtime. In the JAX library, all arrays must have a statically-known shape when JIT compiling functions. On the one hand, this allows JAX to better reason about the program at compile-time (e.g., can two arrays be broadcasted together?), and its compiler to generate highly-optimized binary code under the hood (this much memory is needed here, this looping pattern will be cache-optimal, etc.). On the other hand, this restriction limits the reuse of compilation outputs, and to some degree, the expressibility of the programming language.

Ultimately, this results in the compiler requiring scale parameters to be available as constants in the program. In the case of Shor's, the main problem scale is determined by the number to factor, $N$, or more precisely the number of bits in $N$. This number determines several program parameters, and consequently shape information, like the number of qubits to allocate, the number of loop iterations, and the number of measurement results to collect on the estimation wire.

The restriction to static shapes persists in the presented implementation, leading to recompilation whenever the number of bits in $N$ changes. Even so, avoiding re-compilation for different values of $a$ is used to enhance efficiency during execution. In the future, a fully scale-invariant implementation could be provided by building out the dynamic shape capabilities in Catalyst and JAX. These are currently only available in an experimental state \cite{dynamic-shapes}.

\section{Results}
\label{sec:results}

\begin{figure}[htbp]
\centerline{\includegraphics[width=\columnwidth]{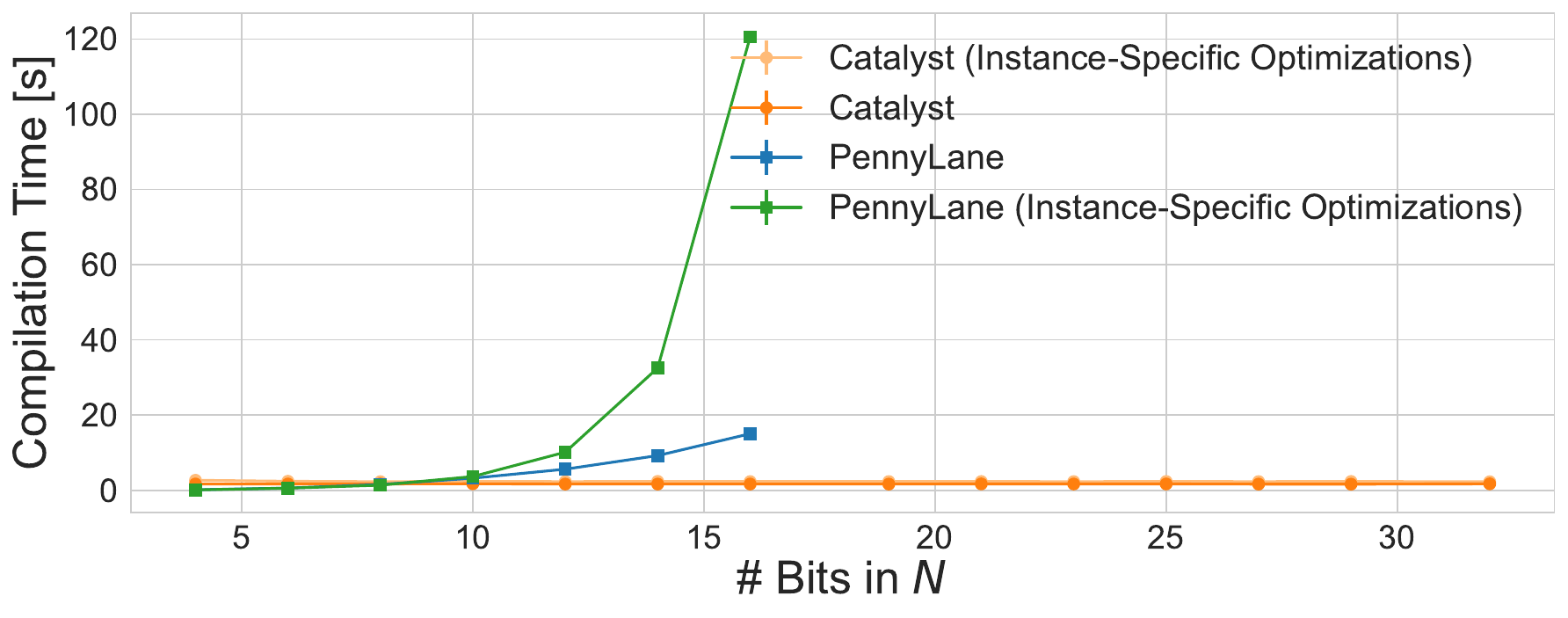}}
\caption[]{Compilation time versus bit width of $N$. Catalyst’s QJIT-compiled implementation exhibits constant compilation time regardless of problem size, in contrast to the exponential scaling observed in standard implementations done with PennyLane. Both implementations are shown with and without the instance-specific optimizations discussed in Sec.~\ref{subsec:a-specific-opts}.}
\label{fig:comp-time-benchmarks}
\end{figure}

\begin{figure}[htbp]
\centerline{\includegraphics[width=\columnwidth]{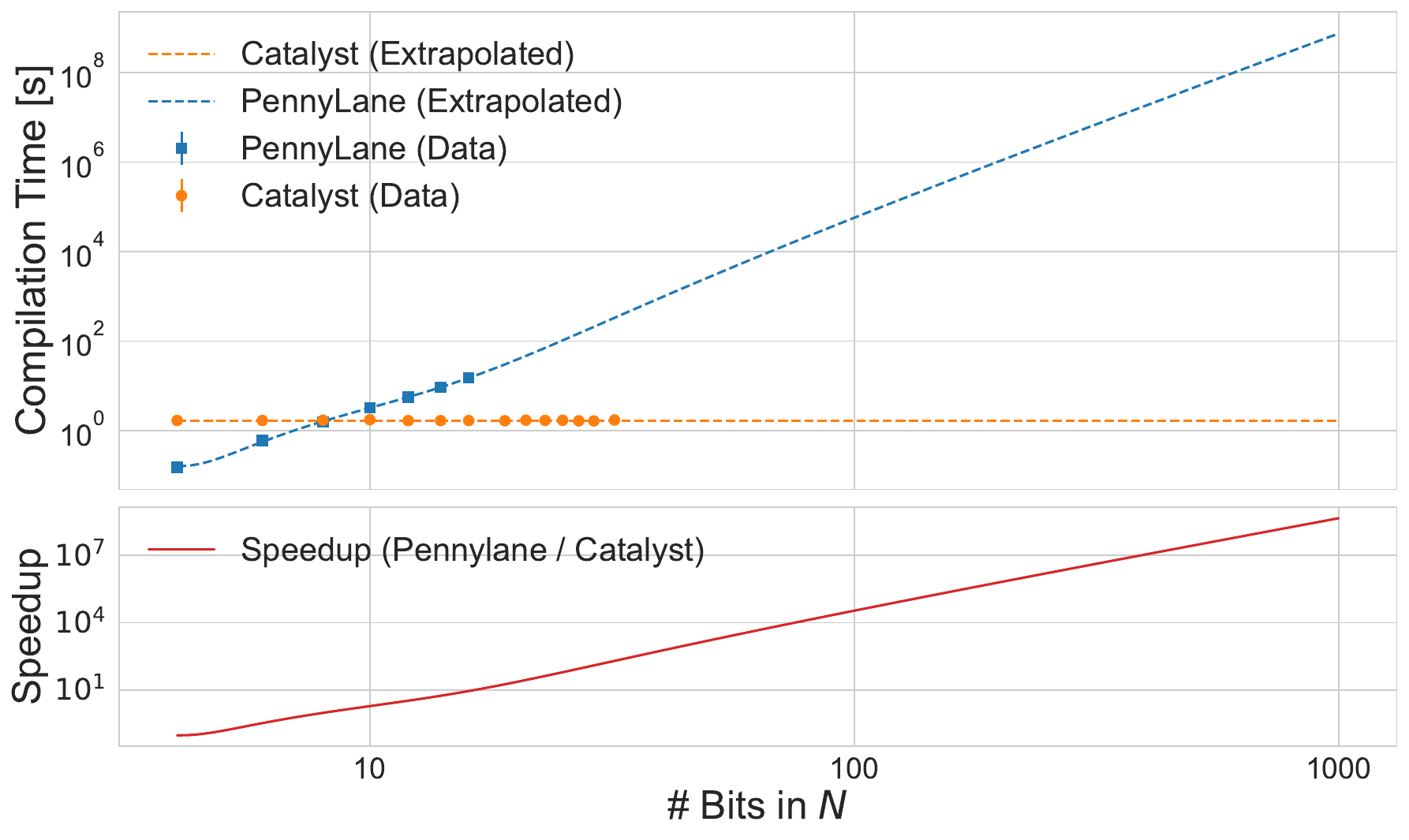}}
\caption[]{Extrapolated compilation times for values of $N$ up to 1000 bits, based on trends from Fig.~\ref{fig:comp-time-benchmarks}. The Catalyst implementation maintains a constant compilation time, while the PennyLane implementation exhibits polynomial growth, fitted with a degree-four curve.}
\label{fig:extrap-comp-time}
\end{figure}

\begin{figure}[htbp]
\centerline{\includegraphics[width=\columnwidth]{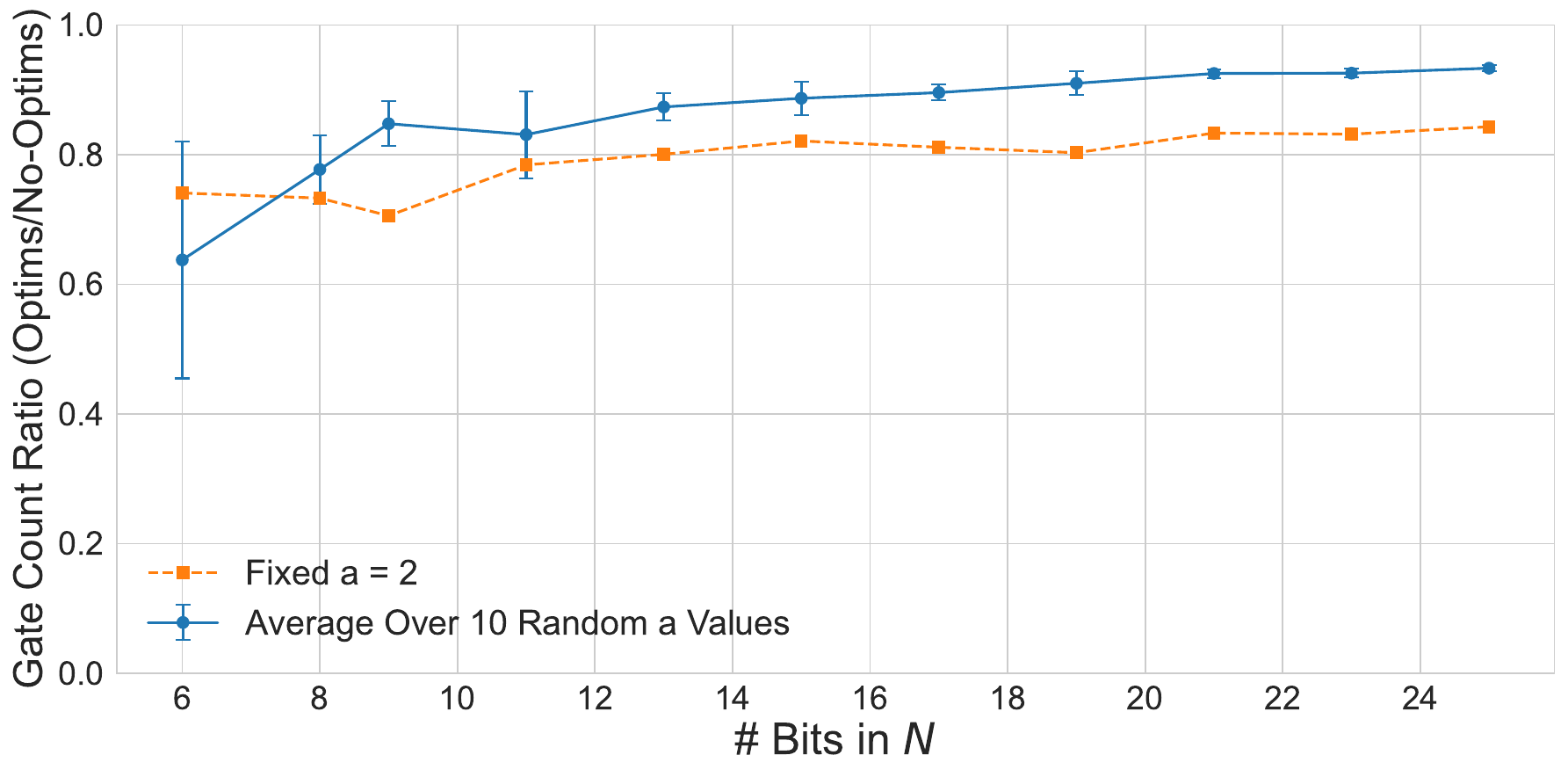}}
\caption[]{Ratio of quantum gate counts for QJIT-compiled Shor's algorithm with and without instance-specific optimizations. The optimized implementation shows a consistent reduction in gate count ranging from 10-25$\%$ for up to 25-bit $N$, averaged over 10 random choices of $a$. A separate line shows results for a fixed $a=2$, which exhibits larger gate count reductions due to the nature of the instance-specific optimizations discussed in Sec.~\ref{subsec:a-specific-opts}.}
\label{fig:gate_count_ratios}
\end{figure}

To evaluate the performance benefits of moving to a scalable QJIT-compiled version, we benchmarked the compilation time for both Catalyst and PennyLane implementations. 
All benchmarking was performed on an Apple M2 Air (16GB RAM). 
Catalyst was tested up to 32-bit values of $N$, while PennyLane was limited to 16-bit $N$ due to increasing resource demands.  
All compilation times reported were averaged over three runs per bit width. The results, shown in Fig.~\ref{fig:comp-time-benchmarks}, include both optimized and unoptimized versions of each implementation, reflecting the impact of the instance-specific optimizations discussed in Section~\ref{subsec:a-specific-opts}. Notably, the QJIT compilation time for Catalyst remains essentially constant at approximately 2.3 seconds for the optimized version, even as the bit width increases to 32. Fig.~\ref{fig:extrap-comp-time} extrapolates these results to larger problem sizes, illustrating that the scalable Catalyst implementation could yield more than a 100-million-fold speedup in compilation time at 1000-bit $N$ compared to traditional approaches. 

The Catalyst implementation with instance-specific optimizations incurs a slightly higher average compilation time of 2.3 seconds, compared to 1.75 seconds for the unoptimized version. However, this overhead is offset by reductions in quantum resource requirements. To quantify these savings, the ratio of total gate count (sum of 1-, 2-, and 3-qubit gates) for values of $N$ up to 25 bits was computed, averaging over 10 random choices of $a$ at each $N$. As shown in Fig.~\ref{fig:gate_count_ratios}, the optimized implementation achieves gate count reductions in the $10$-$25\%$ range, on average.
The $a=2$ case is also plotted, to demonstrate the greater reduction afforded by smaller $a$, which require fewer doubly-controlled additions  early in the QPE procedure.

\section{Conclusions} 
\label{sec:conclusions}

In summary, this work demonstrates how a full hybrid classical-quantum  implementation of Shor's factoring algorithm can be compiled using QJIT compilation. The compilation time is independent of the number being factored, and does not require recompilation for numbers with the same bit width. The techniques are thus scalable out to real-world problem instances. 

The existing implementation has some limitations. Recompilation does occur when changing the bit width due to restrictions in JAX. There are also some utility functions limited to 32- and 64-bit integers (e.g., JAX NumPy's \texttt{unpackbits}). Neither limitation is  insurmountable. A prototype workaround for the first is already supported experimentally in Catalyst; the second is a matter of implementing the utility functions to work with larger bit widths, which is not expected to fundamentally change the scaling of compilation time.

This work also exposes interesting future directions. In particular, a significant amount of manual effort was required to implement the quantum circuits and the subroutine optimizations. Future work should explore how  dynamic circuit optimization at higher abstraction levels can be performed automatically by the QJIT compiler. 

\section*{Acknowledgment}

ODM acknowledges financial support from NSERC (ALLRP 576833-22), the Canada Research Chairs program, and the Department of Electrical and Computer Engineering at UBC. All authors thank the Catalyst and PennyLane teams at Xanadu for useful discussions, and their contribution of features to Catalyst that made this implementation possible. 

\bibliographystyle{ieeetr}
\bibliography{main.bib}

\end{document}